### *In vivo* impact on rabbit subchondral bone of viscosupplementation with a hyaluronic acid antioxidant conjugate.

## R. RIEGER[1,2,*], S. KADERLI[3], C. BOULOCHER[1,4]

[1] Université de Lyon, VetAgro Sup, UPSP ICE 2021.A104, 69280 Marcy-l'Etoile, France
[2] Université de Lyon, École Centrale de Lyon, 69134 Ecully, France
[3] School of Pharmaceutical Sciences, University of Geneva and University of Lausanne, Quai Ernest-Ansermet 30, 1211 Geneva, Switzerland
[4] UniLaSalle Polytechnic Institute, Veterinary College, Campus of Rouen, 76130 Mont Saint Aignan, France

\* Corresponding Author: romain.rieger@ec-lyon.fr (R. Rieger)

Authors:
caroline.boulocher@unilasalle.fr (C. Boulocher)
sema.kaderli@gmail.com (S. Kaderli)
romain.rieger@ec-lyon.fr (R. Rieger)

## Abstract

**Objective**: This study aimed to assess the effects of an antioxidant-conjugated Hyaluronic Acid (HA), specifically HA-4-aminoresorcinol (HA4AR), on articular cartilage and subchondral bone in osteoarthritis (OA). We conducted a comparative analysis between HA4AR and a commercially available high molecular weight HA formulation in a rabbit model of OA.

**Materials and Methods**: Eighteen rabbits underwent unilateral anterior cruciate ligament transection (ACLT) and were divided into three groups of six: Saline-group, HA-group, and HA4AR-group, based on the type of intra-articular injection received. Additionally, eight non-operated contralateral knees served as reference points (Contralateral-group). Six weeks post-surgery, iodine-enhanced micro-computed tomography imaging was used to evaluate articular cartilage volume and thickness, and to assess subchondral bone microarchitecture and mineral density.

**Results**: Cartilage thickness in both the HA and HA4AR groups was comparable to that of the Contralateral group. Notably, there was a significant reduction in subchondral bone plate tissue mineral density in the HA-group when compared to both the HA4AR and Saline groups ($p<0.05$). However, no significant differences in trabecular subchondral bone microarchitectural parameters and mineral densities were observed between the HA4AR-group and the Saline-group. When compared to the Contralateral, Saline, and HA4AR groups, the HA-group exhibited a marked decrease in subchondral bone plate tissue mineral density ($p<0.05$). Additionally, a significant reduction in trabecular bone volume fraction was noted in the HA-group compared to the Contralateral-group ($p<0.05$).

**Conclusions**: The HA-4AR hydrogel demonstrated significant preservation of subchondral bone plate tissue mineral density compared to HA alone, while other bone microarchitectural parameters remained largely unchanged. These findings indicate that HA-4AR may provide enhanced osteoprotective benefits in the treatment of osteoarthritis.

**Keywords**

# 1. Introduction

The degradation of cartilage, a hallmark event in osteoarthritis (OA), is increasingly recognized alongside alterations in subchondral bone as a potential contributory factor to OA onset and progression [1, 2]. Despite advancements, the intricate mechanisms governing subchondral bone remodeling in OA, along with its temporal dynamics, remain incompletely elucidated [3, 4]. Notably, early OA stages manifest heightened bone resorption succeeded by bone accretion, culminating in subchondral bone sclerosis. This has instigated inquiries into therapeutic modalities targeting subchondral bone remodeling [5]. Although intra-articular hyaluronic acid (HA) injections primarily aim to reinstate joint metabolic and rheological equilibrium [6], emerging evidence suggests an interaction with subchondral bone physiology [7–9]. However, the precise effects of HA on bone are obscured, with limited investigations into the ramifications of intra-articular HA injection on pivotal factors governing subchondral bone microarchitecture and strength [10–12].

In the pursuit of prolonging HA residence within the joint space while mitigating potential adverse effects from recurrent intra-articular injections, a strategy involves shielding HA from oxidative degradation through its conjugation with antioxidants such as Mannitol or Sorbitol [13–20]. *In vivo* research, specifically a standardized patient follow-up study, has revealed that the intra-articular administration of a combination of cross-linked hyaluronic acid (HA) and Mannitol significantly mitigates symptoms associated with knee osteoarthritis [15]. Additionally, evidence from a singular animal study suggests that intra-articular injections of HA combined with antioxidants may offer protective effects against the degradation of cartilage [14]. *In vitro* studies converge on the notion that antioxidants such as sorbitol, mannitol, and glutathione, when combined with HA, offer a robust defense against oxidative stress, particularly from reactive oxygen species that are known to degrade HA rapidly [16–18, 20]. Further, this synergistic blend has been shown to not only preserve the rheological properties of HA under oxidative conditions but also to extend its intra-articular residence time, thereby improving viscosupplementation efficacy. Moreover, the cellular and molecular mechanisms illuminated by these studies reveal a promising pathway where such combinations can mitigate oxidative stress, inflammation, and catabolism in joint tissues, underscoring the need for subsequent clinical trials to validate these findings in osteoarthritis management. However, a conspicuous dearth persists in quantitative *in vivo* microarchitectural data concerning the impact of viscosupplementation with HA-antioxidant conjugates on the subchondral bone.

# 2. Materials and methods

## 2.1 Animal model

The experimental procedures involving rabbits were conducted at Centre Lago, Vonnas, France, and implemented in BIOVIVO at the Claude Bourgelat Institute, Lyon, France, with ethical approval granted by the relevant committee and strict adherence to European regulations and ARRIVE (Animal Research: Reporting of In Vivo Experiments) guidelines. A comprehensive ARRIVE checklist is available in the Supplementary Material. We utilized eighteen healthy adult male white New Zealand rabbits, each aged 5 months and weighing approximately 3.68±0.18 kg. Following a two-week acclimation and quarantine period in separate enclosures, we induced experimental osteoarthritis (OA) through unilateral Anterior Cruciate Ligament Transection (ACLT). This surgical procedure was performed by a qualified veterinary surgeon.

Prior to surgery, rabbits received subcutaneous injections, including 30 mg/kg of Borgal® (sulfadoxine and trimethoprim) twice daily, 0.1 mg/kg of morphine, and 0.4 mg/kg of Meloxidyl® (meloxicam). Deep anesthesia was induced through an intramuscular injection of 40 mg/kg Ketamine 1000® and 80 ml/kg Domitor® (medetomidine), maintained by administering 1-3.5% isoflurane through endotracheal intubation.

Following careful shaving and disinfection with Vetedine® (povidone iodine) soap and solution, the surgical procedure was conducted on the left knee using a lateral approach [21], while the right knee remained untouched. Confirmation of complete ACL rupture was achieved by assessing the anterior drawer sign, involving manual induction of horizontal dislocation before sealing the articular capsule. The operated leg was not immobilized, permitting rabbits to move freely in their individual enclosures post-surgery.

## 2.2 Post-surgery care

Following the surgical interventions, specific postoperative measures were implemented to ensure the well-being of the rabbits. They received subcutaneous injections of 0.01 mg/kg buprenorphine twice daily for four days to effectively manage pain. Additionally, a regimen of 0.5 mg/kg Emeprid® (metoclopramide) for three days, 15 mg/kg Borgal® twice daily for nine days, and one daily capsule of Feligastryl® (eserine) for three days was administered to minimize the risk of obstipation. Cothivet® spray was applied to the surgical wound for six days post-procedure. Throughout the recovery period, veterinary professionals closely monitored the rabbits, conducting thorough clinical check-ups every other day. This comprehensive postoperative care protocol not only alleviated pain but also prevented lameness, ensuring the complete recovery of all rabbits from the surgical procedures.

## 2.3 Formulation administration

The personnel responsible for injection administration and assessments were blinded to the specific formulations employed. The 18 rabbits were randomly allocated to one of three groups, each comprising 6 rabbits. Within each group, the left knee subjected to ACLT received an intra-articular injection of 0.2 ml of one of the following substances: a saline solution, a commercial HA formulation (Ostenil®, TRB Chemedica, Switzerland), or HA-4AR [13, 14]. These groups are subsequently denoted as the Saline-group, HA-group, and HA4AR-group. Additionally, eight non-operated right knees were randomly chosen from these groups to serve as non-operated reference points, collectively referred to as the Contralateral-group (n=8). Intra-articular injections were administered at weeks 1, 2, 3, 4, and 5 post-ACLT procedure following a brief period of anesthesia (using 40 mg/kg of Ketamine 1000® and 80 µg/kg of Domitor®) and thorough disinfection with Vetedine® soap and solution.

## 2.4 Grading of osteoarthritis and micro-computed tomography imaging of the subchondral bone

After a 6-week observation period, the rabbits were humanely euthanized through intravascular administration of 1 ml/kg Dolethal® (pentobarbital) following chemical immobilization via intramuscular injection of 40 mg/kg Ketamine 1000® and 80 mg/kg Domitor®. Rigorous dissection of the knees ensued, and the proximal part of the tibia was sectioned using a saw. The assessment of cartilage degradation and the identification of osteophytes were conducted utilizing a macroscopic grading system established by Laverty *et al*. [22]. To determine the osteoarthritis (OA) stage for each tissue, an average OA score was computed based on the grades assigned to the femoral and tibial components before engaging in micro-computed tomography.

Detailed insights into the Equilibrium Partitioning Iodine Contrast Enhanced Micro-Computed Tomography (EPIC-µCT) imaging technique utilizing the eXplore Locus system (General Electric, Fairfield, USA) can be found in the study conducted by Kaderli *et al*. [13]. Image acquisition occurred with an isotropic resolution of 45 µm³ at 80 kV and 450 µA, utilizing a field of view (FOV) with an 80 mm diameter and a depth of 35 mm. Following acquisition and reconstruction, the 16-bit images underwent calibration using a phantom containing hydroxyapatite, water, and air, and were expressed in Hounsfield Units (HU), with air calibrated at 1000 HU and calcified tissues exceeding 100 HU.

## 2.5 Image processing and parameter measurements

A singular operator, blinded to the evaluation, performed the image processing and parameter measurements utilizing MicroView software ABA 2.2 from General Electric in Fairfield, USA. To optimize image quality, an anisotropic filter was applied, and the segmentation process, semi-automatic in nature, relied on the Otsu method [23] to distinguish the subchondral bone plate from the trabecular bone.

Subsequently, the region of interest (ROI) was delineated manually through a contour-based tool within the weight-bearing area of the medial tibial condyle. This involved establishing the X (medio-lateral) and Y (cranio-caudal) axes from the intercondylar area to the lateral edge of the cortical bone and from the intercondylar area to the caudal aspect leading to the medial condyle, respectively. The Z axis (proximo-distal axis) was adapted to each tissue type: for the subchondral bone plate, it extended from the calcified cartilage to the end of the subchondral bone plate, and for the trabecular bone, it spanned from the subchondral bone plate/trabecular bone junction to the end of the epiphyseal line. Approximately 100 slices were marked for the subchondral cortical bone, while around 80 slices were delineated for the trabecular bone. Variations in slice numbers were attributed to the trabecular volume ROI becoming too small at the intersection of the intercondylar area and the epiphyseal line.

The microarchitectural parameters were subsequently computed using the conventional 2D histomorphometric method [24]. These parameters encompassed: (i) mean subchondral bone plate thickness (Pt.Th [mm]) for the subchondral bone plate; (ii) mean trabecular thickness (Tb.Th [mm]), mean trabecular separation (Tb.Sp [mm]), mean trabecular bone volume fraction (Tb.BVF [%]), and mean trabecular bone mineral density (Tb.BMD [mg of mineral content per cc]) for the trabecular bone. Additionally, mean bone tissue mineral density (TMD [mg of mineral content per cc]) was calculated for both the subchondral bone plate (referred to as Pt.TMD) and the trabecular bone (referred to as Tb.TMD). A 3D local thickness measure was obtained for each voxel within the ROI, based on the method proposed by Hildebrand and Rüegsegger [25].

## 2.6 Statistical analysis of microarchitectural and mineral density parameters

R software (R Foundation for Statistical Computing, Vienna, Austria version 3.1.2014-10-26) served as the platform for our statistical analyses. Each group and variable underwent Shapiro-Wilk's and Levene's tests to assess data normality and equality of variances. In cases where normality and equal variances were not confirmed for specific variables, non-parametric tests were implemented.

Initiating the analysis, the Kruskal-Wallis ANOVA test, set at a significance level ($\alpha$) of 0.05, was deployed. Subsequently, when statistical significance was detected, post hoc analysis ensued, employing the Mann-Whitney-Wilcoxon rank test ($\alpha$=0.05).

The primary comparisons encompassed evaluating the impact of ACLT on subchondral bone at the 6-week post-surgery stage, with the Saline-group, HA-group, and HA4AR-group compared to the Contralateral-group. Next, the effects of viscosupplementation on subchondral bone were explored by comparing the HA-group and HA4AR-group to the Saline-group. Finally, to investigate the combined effect of the HA-antioxidant conjugate and HA, the HA4AR-group was compared to the HA-group.

Significance levels were denoted as follows: * for p-values < 0.05, ** for p-values < 0.01, and *** for p-values < 0.001. The degrees of freedom (E) in the analysis of variance were calculated using the formula: E = Total number of animals – Total number of groups. With 18 rabbits and four groups (Contralateral, Saline, HA, and HA4AR groups), E equaled 18 – 4 = 14, falling within the recommended range of 10 to 20 for E [29]. Thus, the sample size in this study appears appropriate for assessing statistical significance.

## 2.7 Hypothesis and Rationale

Drawing upon the existing literature that underscores the impact of antioxidants on chondrocytes, with a particular focus on the chondroprotective role of an antioxidant-conjugated hyaluronic acid (HA) formulation, HA-4-aminoresorcinol (HA-4AR), as formulated by Kaderli *et al.* [14], this study delves into the potential osteoprotective effect of HA-4AR in mitigating accelerated bone remodeling in early-stage osteoarthritis. Our hypothesis is based on the premise that the observed benefits in chondrocytes might also extend to bone cells, thereby offering a comprehensive protective mechanism within the joint. To investigate this, we compare the impact of HA-4AR with a standard high molecular weight HA formulation on the microarchitecture and mineral density of subchondral bone.

Our evaluation encompasses a detailed analysis of microarchitectural features of the subchondral bone, including mean subchondral bone plate thickness (Pt.Th [mm]); mean trabecular thickness (Tb.Th [mm]), separation (Tb.Sp [mm]), bone volume fraction (Tb.BVF [%]), and mineral density (Tb.BMD [mg of mineral content per cc]); along with the mean bone tissue mineral density (TMD [mg of mineral content per cc]) for both the subchondral bone plate (Pt.TMD) and trabecular bone (Tb.TMD). These parameters are crucial for assessing the integrity of subchondral bone components. The evaluations were carried out during the early phase of osteoarthritis, induced post-anterior cruciate ligament transection (ACLT) in rabbits, a well-established model for studying the pathogenesis and efficacy of OA treatments [21, 26–28]. A 6-week time point was selected to capture early bone changes in the ACLT rabbit model of OA, consistent with the initial phases of OA development within the knee joint [26, 29] and in line with previous research [30, 31] that characterize osteoarticular lesions.

# 3 Results

## 3.1 Assessment of Osteoarthritis Lesions

### *Macroscopic Grading of osteoarthritis*
All surgically treated groups exhibited macroscopic osteoarthritis (OA) lesions, with mean OA scores of $1.8 \pm 0.72$ for the Saline-group, $2.14 \pm 0.62$ for the HA-group, and $1.63 \pm 0.64$ for the HA4AR-group (data not shown). In contrast, the contralateral knees showed no macroscopic signs of OA.

### *Micro-Tomography Imaging and Quantitative analysis*
Representative micro-tomography images of tibial plates from the HA4AR-group and Contralateral-group are presented in Figure 1. No discernible gross differences in subchondral bone microarchitecture were observed between these groups on imaging.

To further investigate the local impact of viscosupplementation, we conducted quantitative analyses on micro-tomography images to assess subchondral bone microarchitectural parameters and mineral density for all groups, as displayed in Table 1.

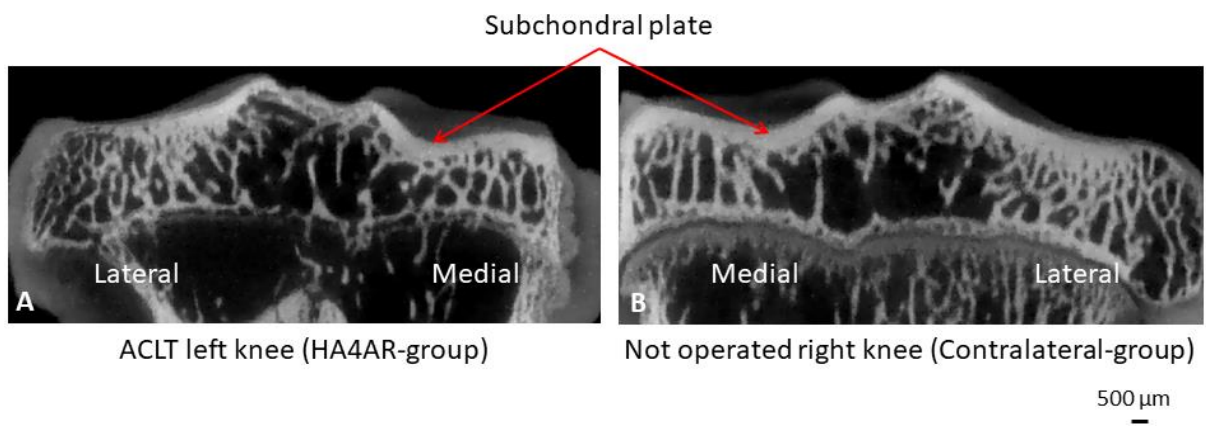

Figure 1. Typical ex vivo microtomography images of tibial plate (6 weeks after ACLT). A/ HA4AR-group (left knee); B/ Contralateral-group (right knee).

Table 1: Summary of microstructure parameters

| Parameter | Contralateral-group | Saline-group | HA-group | HA4AR-group |
|---|---|---|---|---|
| Ct.Th [mm] | 0.66±0.03 | 0.57±0.04 [(a)] | 0.64±0.11 | 0.62±0.08 |
| Pt.Th [mm] | 0.4±0.05 | 0.36±0.05 | 0.39±0.07 | 0.35±0.04 |
| Tb.Th [mm] | 0.16±0.02 | 0.14±0.01 | 0.14±0.02 | 0.13±0.02 |
| Tb.Sp [mm] | 0.27±0.05 | 0.29±0.05 | 0.33±0.05 | 0.29±0.07 |
| Tb.BVF [ %] | 0.49±0.06 | 0.44±0.03 | 0.4±0.03 [(a)] | 0.43±0.06 |
| Tb.BMD [mg/cc] | 387.79±46.57 | 319.5±28.48 [(a)] | 284.26±31.82 [(a)] | 309.75±43.73 [(a)] |
| Tb.TMD [mg/cc] | 553.21±32.9 | 497.15±26.81 [(a)] | 473.74±33.06 [(a)] | 483.96±36.93 [(a)] |
| Pt.TMD [mg/cc] | 622.78±22.5 | 616.15±31.94 | 561.55±28.23 [(a,b)] | 597.68±30.52 [(c)] |

Table 1. Articular cartilage thickness, bone microarchitectural parameters and bone mineral densities of medial tibial condyles: cartilage (Ct), subchondral bone plate (Pt) and trabecular bone (Tb) at 6 weeks post-surgery and their statistical significance (p<0.05) when compared to Contralateral-group: a, Saline-group: b or HA-group: c.

*Comparison of Lateral and Medial Tibial Condyle Parameters*

Articular cartilage thickness (C.Th), bone microarchitectural parameters, and bone mineral density on the lateral tibial condyle exhibited similar significant alterations to those observed in the medial tibial condyle (data not shown).

## 3.2 Comparison to Contralateral-group

Figure 2 illustrates cartilage thickness (C.Th) and bone microarchitectural parameters. The mean value of C.Th (Figure 2A) was significantly lower in the Saline-group than in the Contralateral-group (p<0.01). No significant differences in C.Th mean values were observed in the HA-group and HA4AR-group compared to the Contralateral-group.

Both Saline and HA groups exhibited decreased mean values in Tb.Th and Tb.BVF and increased mean values in Tb.Sp compared to the Contralateral-group (Figure 2C, D, E), although not statistically significant except for Tb.BVF in the HA-group (p<0.01). The HA4AR-group showed decreased mean values in Pt.Th, Tb.Th, Tb.BVF, and increased mean values in Tb.Sp compared to the Contralateral-group (Figure 2B, C, D, E), although not statistically significant.

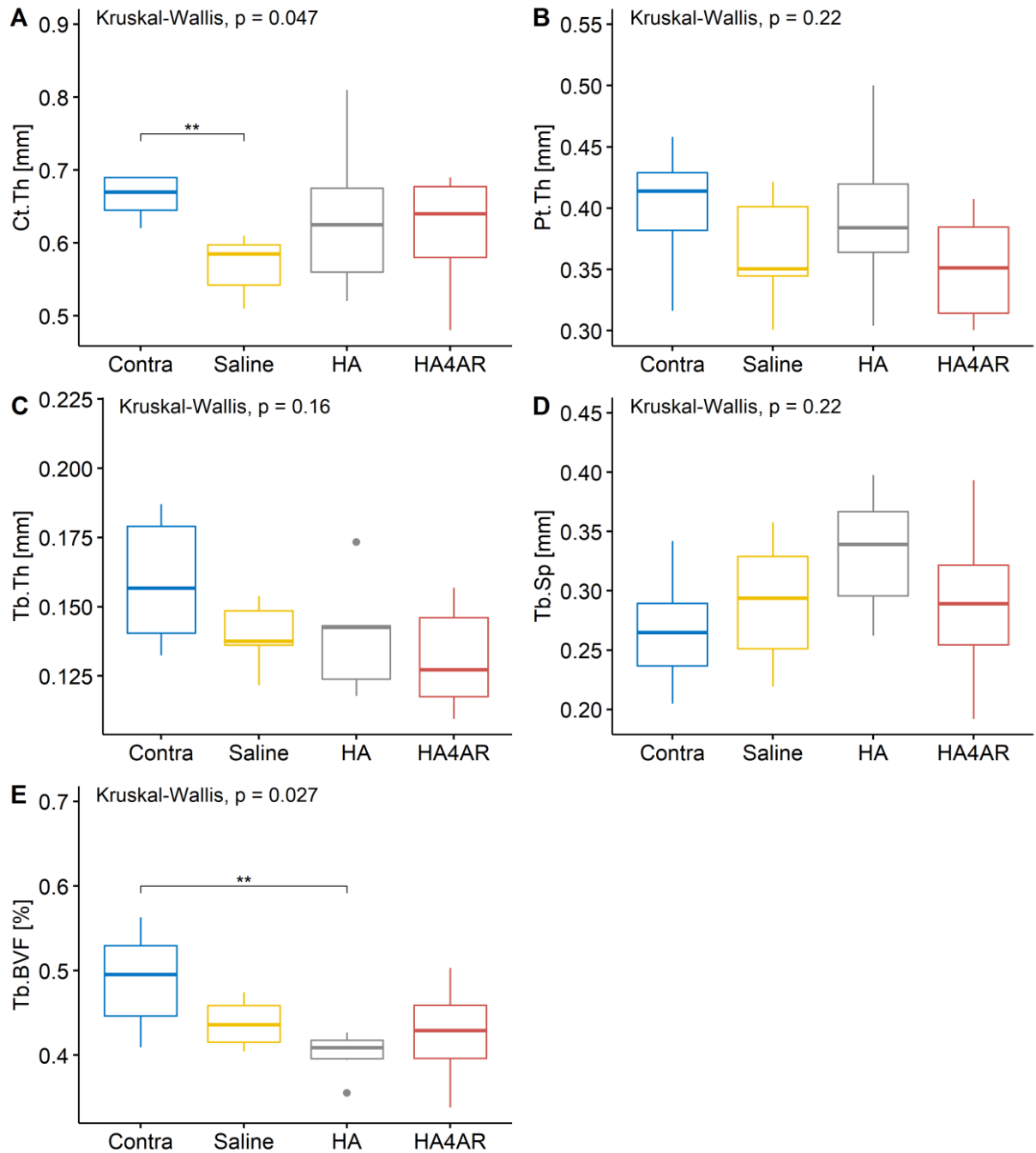

Figure 2. Cartilage thickness and microarchitectural parameters (mean ± SD) in Contralateral, Saline, HA and HA4AR groups. A/ cartilage thickness (Ct.Th); B/ subchondral bone plate thickness (Pt.Th); C/ trabecular thickness (Tb.Th); D/ trabecular separation (Tb.Sp); E/ trabecular bone volume fraction (Tb.BVF). * : p<0.05, ** : p<0.01 and *** : p<0.001 indicate statistical significance under pairwise Mann-Whitney-Wilcoxon Rank test (α=0.05) performed after significant Kruskall-Wallis ANOVA test (α=0.05).

Mean values of bone mineral densities (Tb.BMD, Tb.TMD, Pt.TMD) are presented in Figure 3. In the Saline-group compared to the Contralateral-group, Tb.BMD and Tb.TMD mean values (Figure 3A-B) were significantly decreased (p<0.05 and p<0.01, respectively). In both HA and HA4AR groups compared to the Contralateral-group, Tb.BMD and Tb.TMD mean values (Figure 3A-B) were significantly decreased (p<0.01 and p<0.01, respectively for the HA-group and p<0.05 and p<0.01, respectively for the HA4AR-group). Regarding Pt.TMD in HA and HA4AR groups, mean values were decreased compared to the Contralateral-group (Figure 3C), although not statistically significant.

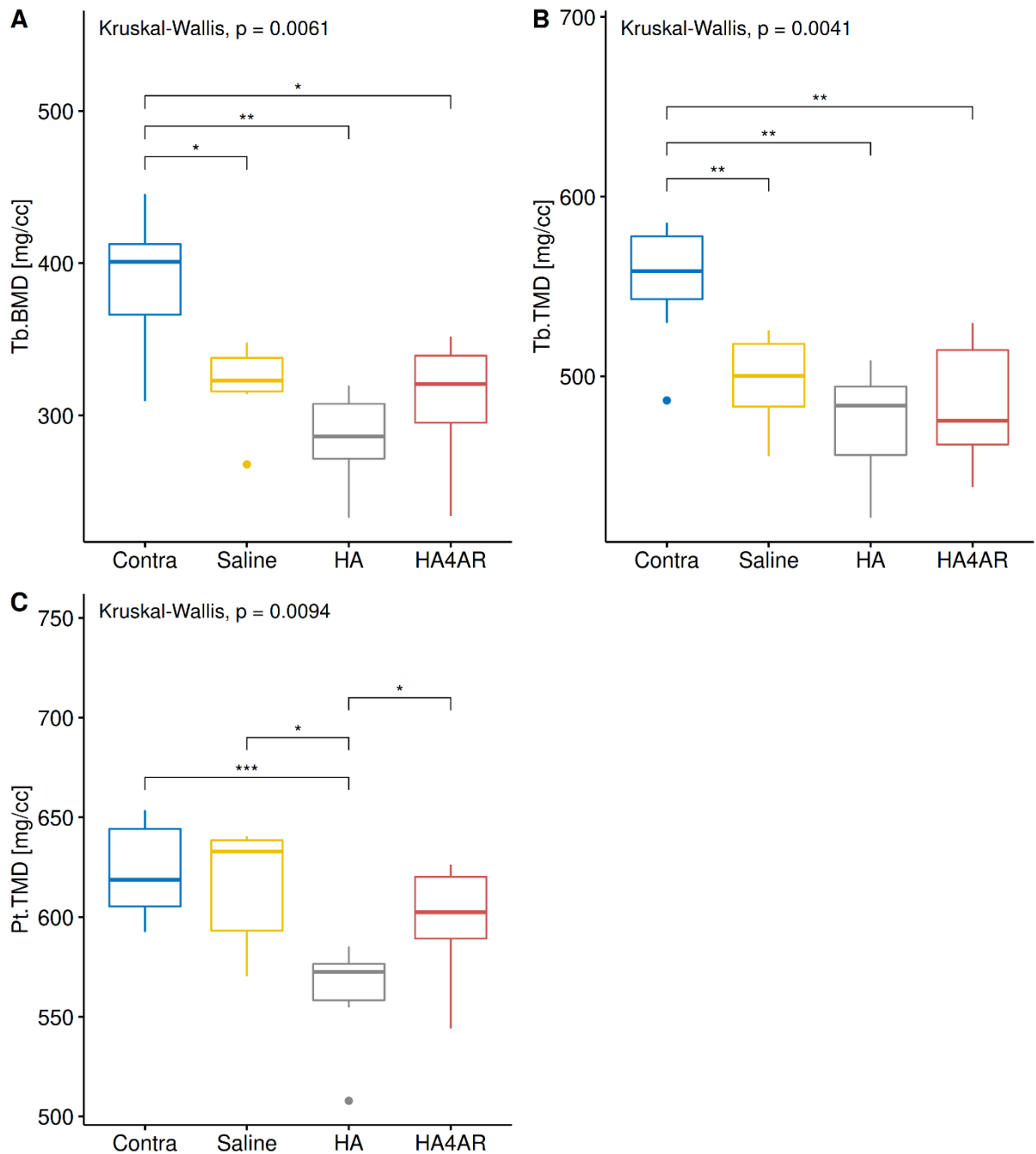

Figure 3. Bone Mineral Density (mean ± SD) in Contralateral, Saline, HA and HA4AR groups. A/ trabecular bone BMD (Tb.BMD), B/ trabecular bone TMD (Tb.TMD); C/ subchondral bone plate TMD (Pt.TMD). * : p<0.05, ** : p<0.01 and *** : p<0.001 indicate statistical significance under pairwise Mann-Whitney-Wilcoxon Rank test (α=0.05) performed after significant Kruskall-Wallis ANOVA test (α=0.05).

### 3.3 Comparison to Saline-group

Figure 2 illustrates the comparison of cartilage thickness and bone microarchitectural parameters. The mean values of C.Th were higher in the HA and HA4AR groups compared to the Saline-group (Figure 2A), although these differences were not statistically significant.

In the HA-group, Pt.Th and Tb.Sp exhibited higher mean values, while Tb.Th and Tb.BVF showed lower mean values compared to the Saline-group (Figure 2B-E), although these differences were not statistically significant.

For the HA4AR-group, only Tb.Th mean value was lower compared to the Saline-group (Figure 2E), although not statistically significant.

Regarding bone mineral densities, in the HA-group compared to the Saline-group, Tb.BMD, Tb.TMD, and Pt.TMD mean values were decreased (Figure 3), although not statistically significant except for Pt.TMD ($p<0.001$). In the HA4AR-group compared to the Saline-group, there were no significant differences for Tb.BMD, Tb.TMD, and Pt.TMD (Figure 3).

### 3.4 Comparison between HA-group and HA4AR-group

C.Th mean values did not differ between the HA and HA4AR groups (Figure 2A).

In the HA4AR-group, Pt.Th and Tb.BVF showed lower mean values, while Tb.Sp showed a higher mean value compared to the HA-group (Figure 2B-E), although these differences were not statistically significant.

Regarding bone mineral densities in the HA4AR-group compared to the HA-group, Tb.BMD, and Pt.TMD showed higher mean values (Figure 3 B, C), although not statistically significant except for Pt.TMD ($p<0.05$).

Articular cartilage thickness, bone microarchitectural parameters, and bone mineral density mean values on the lateral tibial condyle exhibited the same significant changes as the medial tibial condyle (data not shown).

## 4 Discussion

This study sought to assess the impact of an HA-antioxidant conjugate, specifically HA-4AR hydrogel, on subchondral bone during viscosupplementation, compared to a high molecular weight HA formulation [13, 14]. The evaluation focused on subchondral bone microarchitectural parameters, bone mineral density, and cartilage thickness in a rabbit model of osteoarthritis (OA), six weeks post-surgery. While EPIC micro-CT has demonstrated efficacy in distinguishing articular cartilage from calcified bone, its ability to differentiate between articular and calcified cartilage is not explicitly demonstrated in our study [32]. The challenge lies in the similar X-ray attenuation properties shared by calcified cartilage and subchondral bone, posing difficulty in their separation in micro-CT images [33].

Contrary to our initial hypothesis, the results did not affirm the idea that HA-4AR would maintain the overall subchondral bone microarchitecture and bone mineral density. However, we did observe significant preservation of mineral density in the subchondral bone plate tissue in the HA4AR-group.

### Effect of OA induction 6 weeks post-surgery

At the chosen 6-week time point, aligning with the early phase of OA characterized by initial bone resorption [3, 29–31], a notable decrease in cartilage thickness (C.Th) was observed in the Saline group compared to the Contralateral group, confirming OA presence. However, subchondral bone parameters remained largely unchanged. In contrast, the HA and HA4AR groups showed no significant decrease in C.Th compared to the Contralateral group, suggesting that HA may slow down cartilage degradation, an effect not enhanced by HA-4AR, consistent with previous findings of HA's chondroprotective role [7, 10, 34, 35].

### Effect of HA

In the HA-group, a notable reduction was observed in subchondral trabecular bone volume fraction (BVF) and mineral densities (Tb.BMD, Tb.TMD, and Pt.TMD), as depicted in Figure 2E and Figure 3, when compared to the Contralateral-group. This finding aligns with the results of Duan *et al.* [36] , who reported a decrease in BVF with HA treatment compared to a non-treated group, although statistical significance was not attained. The disparity between our study and that of Duan *et al.* [31] could be attributed to differences in time points (42 days in our study vs. 56 days) and the method of OA induction (ACLT in the present study vs. axial tibial loading) potentially resulting in a less severe OA in the latter.

Conversely, our results contrast with those of Shen *et al.* [9], who, in a murine model of joint injury, found no discernible effect of HA-based viscosupplementation. The discrepancy might stem from variations in the animal model, as they employed a single injection compared to the five injections administered in our study. The use of multiple injections is often justified in clinical practice [37] due to the short residency time of HA-based viscosupplements [37, 38]. Additionally, in a rabbit model of OA, Permuy *et al.* [8] did not observe significant subchondral bone loss following HA intra-articular injection. The contrasting results could be attributed to differences in the severity of induction techniques (combining ACLT and meniscectomy vs. ACLT) and the later endpoint used in their study (11 vs. 6 weeks), leading to a more advanced OA and potential recovery of initial subchondral bone microarchitectural parameters and mineral density [3].

Overall, the decline in subchondral bone microarchitectural parameters in the HA-group was linked to demineralization, suggesting subchondral bone resorption, possibly due to the influence of HA on osteoclast activity in OA subchondral bone [34].

### Effect of HA-4AR

In the HA4AR-group, a range of subchondral bone microarchitectural parameters showed a decrease compared to the HA-group, as depicted in Figures 2B-D. This led to a reduced thickness in the subchondral bone plate and an increased trabecular bone volume fraction, though these changes did not reach statistical significance. Importantly, the subchondral bone mineral densities were notably higher in the HA4AR-group than in the HA-group, with a statistically significant difference observed in the subchondral bone plate, as shown in Figure 3C. This increase in bone mineralization across the subchondral bone implies that the HA-antioxidant conjugate, compared to HA alone, may reduce bone resorption. Further support for the superior osteoprotective effect of HA-4-aminoresorcinol (HA4AR) over a commercial HA formulation, such as Ostenil®, comes from a parallel study. This study, which compared Ostenil® with a hybrid hydrogel that combines HA and chitosan [39], demonstrated a significant decrease in bone tissue mineral densities in both the HA and Hybrid groups compared to the Saline group, thereby corroborating the findings of enhanced osteoprotection by HA4AR observed in our study.

### Contralateral-reference and OA assessment

In the current investigation, contralateral knees served as references for evaluating osteoarthritis (OA) gross lesions. X-ray osteophyte scoring and a macroscopic grading system were employed for this purpose, revealing no discernible changes in subchondral bone or cartilage in any of the contralateral knees. The use of the contralateral limb as a reference offers notable advantages in terms of cost and minimizing the number of animals required. Recent studies have adopted this approach [9, 36], and Hintz *et al.* [40]. demonstrated no significant differences in OARSI scores on contralateral knees in a short-term (4 weeks) destabilization of medial meniscus (DMM) OA rat model. Although data are limited regarding the initial signs of OA in subchondral bone changes, a comparison between anterior cruciate ligament transection (ACLT) and DMM OA induction time frames in various rat models [41–43] suggests that DMM leads to more moderate degenerative changes compared to ACLT at 4 weeks post-surgery. However, it is essential to consider that larger animal models generally exhibit a slower onset of OA degradation signs. Therefore, in terms of the time frame for OA changes, the induction of DMM OA in a 4-week rat model might be equivalent to ACLT OA induction in a 6-week rabbit model. Collectively, these considerations support the use of the contralateral limb as a reference rather than a control in our short-term study.

### Limitations of the Study

This study has several limitations that should be acknowledged. Firstly, the resolution of the micro-CT used in this study was 45 µm, which may not have been sufficient to distinguish between calcified cartilage and subchondral bone. Additionally, the relatively short duration of 6 weeks may not capture the long-term effects of HA-4AR on subchondral bone remodeling. The use of only male rabbits also limits the generalizability of the findings to female rabbits or other species. Finally, the study did not include functional assessments of joint health, such as pain or mobility, which could provide more comprehensive insights into the therapeutic effects of HA-4AR.

### Strengths and Novelties

Despite these limitations, the study presents several strengths and novel findings. It is one of the first studies to investigate the osteoprotective effects of an HA-antioxidant conjugate in a rabbit model of OA, providing valuable insights into its potential benefits. The use of a well-established ACLT model ensures the relevance of the findings to human OA. Additionally, the study's comprehensive analysis of both microarchitectural parameters and bone mineral densities offers a detailed understanding of the effects of HA-4AR on subchondral bone. The significant preservation of subchondral bone plate mineral density observed with HA-4AR highlights its potential as a more effective treatment compared to conventional HA formulations.

# 5 Conclusion

In this study, we endeavored to elucidate the comparative impacts of an HA-antioxidant conjugate (HA-4AR) and HA alone on cartilage thickness, subchondral bone microarchitecture, and mineral density. Our investigation, the first of its kind to our knowledge, focused on quantifying the effects of HA-antioxidant conjugate viscosupplementation in a rabbit model of OA. The results underscored a significant preservation of mineral density in the subchondral bone plate tissue when using the HA-4AR hydrogel. While other microarchitectural parameters of the bone did not exhibit substantial modifications, a trend towards preservation was noted in the trabecular bone volume fraction in the HA4AR-group compared to the HA-group, specifically observed six weeks post-ACLT in rabbits.

While the comparison of HA-4AR with saline does not demonstrate a clear advantage of viscosupplementation beyond saline solution, a significant osteoprotective effect becomes apparent when contrasting HA-4AR with HA, especially in terms of Pt.TMD. This contrast draws attention to HA's inherent limitations as a standalone viscosupplement for osteoprotective purposes, limitations that appear to be effectively addressed by the HA-4-aminoresorcinol solution. This revelation advocates for a more comprehensive approach to evaluating viscosupplementation, one that considers the osteoarticular joint in its entirety rather than concentrating exclusively on cartilage and soft tissues.

In conclusion, the study provides valuable contributions to understanding the specific impact of HA-4AR on subchondral bone, particularly its potential in preserving mineral density. These findings augment our comprehension of viscosupplementation's broader effects within the context of OA, potentially guiding future therapeutic strategies.

# Declarations

**a. Clinical trial number**

Not applicable

**b. Ethics approval and consent to participate**

Not applicable

**c. Consent for publication**

Not applicable

**d. Availability of data and materials**

The datasets used and/or analysed during the current study are available from the corresponding author on reasonable request.

**e. Competing interests**

The authors declared no potential conflicts of interest with respect to the research, authorship, and/or publication of this article.

**f. Funding**

The authors received no external financial support for the research, authorship, and/or publication of this article.

**g. Authors' contributions**

Romain Rieger: Conceptualization; Formal analysis; Investigation; Methodology; Project, supervision, Administration; Software; Validation; Visualization; Roles/Writing - original draft; Writing - review & editing.

Sema Kaderli: Data curation, Conceptualization; Resources; Writing - review & editing.

Caroline Boulocher: Conceptualization, methodology, validation, writing - review & editing, supervision, project administration, funding acquisition.

### h.  Acknowledgements

The authors wish to thank BIOVIVO from the Claude Bourgelat Institute, Lyon, France, and Centre Lago, Vonnas, France for the animal care and logistical assistance. We would also like to thank Robert Gurny for giving us the opportunity to further characterize the tissues from his previous work, the LTDS laboratory and the IVTV ANR-10-EQPX-06-01 for providing a working space.